\documentclass[12pt]{article}

\usepackage{tabularx}
\usepackage{scicite}
\usepackage{booktabs}
\usepackage{adjustbox}
\usepackage{graphicx}
\usepackage{float}
\usepackage{caption}

\usepackage{times}

\newenvironment{sciabstract}{%
\begin{quote} \bf}
{\end{quote}}

\newcounter{lastnote}

\title{Observation of odd-parity superconductivity in UTe$_2$}

\author
{Zixuan Li$^{1}$,  Camilla M. Moir$^{2}$, Nathan J. McKee$^{1}$, Eric Lee-Wong$^{3}$, \\
Ryan E. Baumbach$^{4, 5}$,  M. Brian Maple$^{2\ast}$, Ying Liu$^{1\ast}$\\
\\
\normalsize{$^{1}$Department of Physics and Materials Research Institute, The Pennsylvania State University,}\\
\normalsize{University Park, PA 16802, U.S.A.}\\
\normalsize{$^{2}$Department of Physics, University of California San Diego, La Jolla, CA 92093, U.S.A.}\\
\normalsize{$^{3}$Department of NanoEngineering, University of California, San Diego, CA 92093, U.S.A.}\\
\normalsize{$^{4}$National High Magnetic Field Laboratory, Florida State University, Tallahassee, FL 32310, U.S.A.}\\ 
\normalsize{$^{5}$Department of Physics, Florida State University, Tallahassee, FL 32306, U.S.A.}\\
\\
\normalsize{$^\ast$Corresponding authors. E-mail: mbmaple@ucsd.edu (M.B.M.); yxl15@psu.edu (Y.L.)}
}

\date{\today}

\begin{document} 
\baselineskip24pt
\maketitle 
\hyphenpenalty=10000
\exhyphenpenalty=10000
\begin{sciabstract}
Symmetry properties of the order parameter are among the most fundamental characteristics of a superconductor. The pairing symmetry of recently discovered heavy fermion superconductor UTe$_2$ featuring an exceedingly large upper critical field has attracted a great deal of attention. Even though it is widely believed that UTe$_2$ possesses an odd-parity, spin-triplet pairing symmetry, direct evidence for it is lacking, especially at zero or low magnetic fields. We report here the selection-rule results of Josephson coupling between In, an $s$-wave superconductor, and UTe$_2$. The orientation dependence of the Josephson coupling suggests very strongly that UTe$_2$ possess an odd-parity pairing state of B$_{1u}$ in zero magnetic fields. We also report the formation of Andreev surface bound states on the (1-10) surface of UTe$_2$.
\end{sciabstract}

The order parameter (OP) of a superconductor, or the wave function (WF) of the Cooper pairs, consists of an orbital and a spin part. The search for superconductors featuring an OP different from that used in the original Bardeen-Cooper-Schrieffer theory for superconductivity\cite{BCS_1957}, a spin-singlet, $s$-wave one, has been an important direction for superconductivity research\cite{Mineev_1999,Leggett_2006}. UTe$_2$, a newly discovered heavy fermion superconductor featuring a superconducting transition temperature (T$_c$) up to 2K, is widely believed to feature odd-parity, spin-triplet pairing\cite{ran_discovery_2019,Ran_field-boosted_2019} due to the observation of an extremely high upper critical field along the $b$ axis, $H_{c_2,b} = 20~T$, for a low T$_c$ superconductor, which greatly exceeds the Pauli paramagnetic limit for spin-singlet superconductors\cite{Aoki_specific_heat_2019,Aoki_Review_2022}. Above $H_{c_2,b}$, spectacular re-entrant superconductivity was observed\cite{Ran_field-boosted_2019}. A rich phase diagram was found when the field is tilted away from the $b$ axis. In particular, when the field is tilted towards the $c$ axis, re-entrant superconductivity was found unexpectedly within a range of tilting angle and field strength\cite{Ran_field-boosted_2019}. As the field strength increased further, a metamagnetic field transition into a ferromagnetic phase was observed\cite{knebel_reentrant_2019}. On one hand, the presence of metamagnetism suggest that the material is near ferromagnetic instability. On the other hand, neutron scattering studies revealed that the dominant magnetic fluctuations are incommensurate\cite{Butch_MagneticCorrelation_2022,knafo_neutron_2021}. In any case, complex magnetic fluctuations were shown theoretically to lead to spin-triplet pairing\cite{mineev_fluctuations_2014,ishizuka_Anderson_model_2021}.

Whether UTe$_2$ is indeed a spin-triplet superconductor is not yet clear, at least not in zero or low applied magnetic fields. UTe$_2$ is a multiband superconductor featuring strong anisotropy in the upper critical field\cite{Aoki_specific_heat_2019}. It was known theoretically that for multiband spin-triplet superconductors the magnitude of the upper critical field is a complicated question\cite{LeggettLiu_2021}. The large spin-orbital coupling (SOC) expected in UTe$_2$ leads to further complications in linking the high upper critical fields to spin-triplet pairing. A small drop in the NMR Knight shift with the field along the $b$ axis was found\cite{Fujibayashi_KnightShift_2022}. Such a small decrease can be explained by singlet pairing in the presence of SOC as opposed to spin-triplet pairing\cite{LeggettLiu_2021}. The absence of the Hebel-Slichter coherence peak in the 1${/}T_1$ data\cite{Nakamine_NMR_2019} where 1${/}T_1$ is nuclear spin-lattice relaxation rate, can be explained equally well in a non-$s$-wave spin-singlet pairing picture. Thus experiments aiming at determining the symmetry of the OP of UTe$_2$ without limitations described above are urgently needed.  

\begin{figure}[H]
\centering
\includegraphics[scale=1]{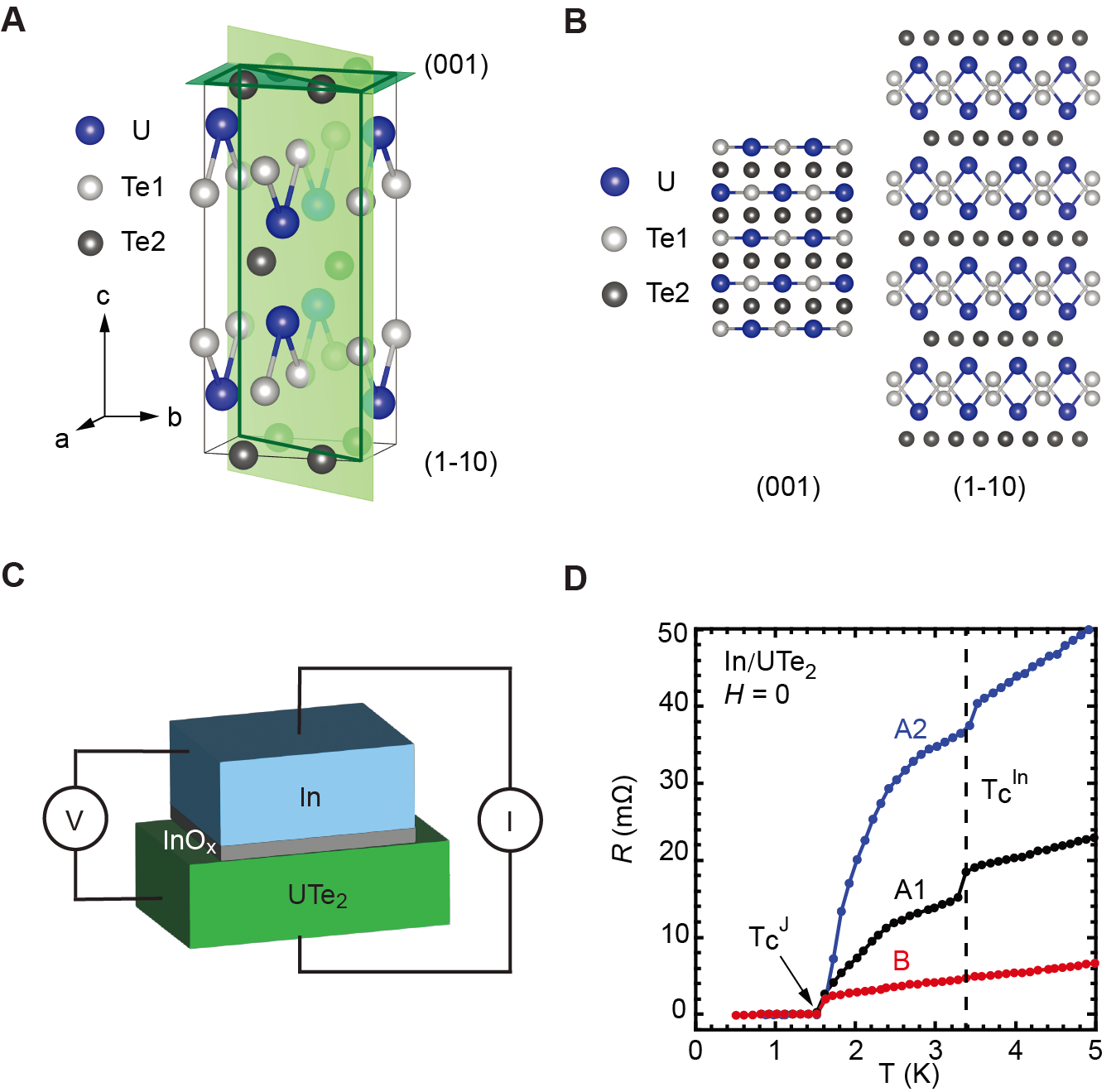}
\phantomcaption
\label{fig:1}
\end{figure}
\noindent {\bf Fig. 1. Structure of UTe$_2$ and Josephson junction.} 
(A) Crystal structure of UTe$_2$ featuring space group \textit{Immm} (\#71) with orthorhombic point group of D$_{2h}$. (B) Schematics of crystalline surfaces of (001) and (1-10) showing very different atomic arrangements. (C) Schematic for an In/UTe$_2$ Josephson junction prepared on an as-grown flat surface of UTe$_2$. (D) Resistance ($R$) $vs$. temperature ($T$) curves in zero magnetic fields for Samples A1, A2, and B. Drops at 1.6 K and 3.4 K correspond to the T$_c$'s for UTe$_2$ and In, respectively. A small feature was also seen at 3.4 K in Sample B, the T$_c$ of In.  

For crystalline superconductors featuring an inversion symmetry, the OP must be of either even-parity, spin-singlet or odd-parity, spin-triplet if the finite-momentum or odd-frequency pairing is excluded\cite{LeggettLiu_2021}. According to the Volovik-Gor'kov theory\cite{VolovikGor'kov_1985}, the pairing symmetries can be specified by the basis functions of the irreducible representations (irreps) of the symmetry group of the normal state, G$\times$T$\times$U(1), where G is the point-group, T the time-reversal, and U(1) the gauge symmetries, respectively. UTe$_2$ has a body-centered orthorhombic crystal structure (Fig. $\ref{fig:1}$A). Various crystalline surfaces may be obtained by cleaving (Figs. $\ref{fig:1}$B). Its space group is \textit{Immm} with the point group $D_{2h}$. Pairing states allowed by this group are listed in Table 1. In the spin-triplet case, the orbital and spin rotations are locked together due to the strong SOC expected in UTe$_2$. 

\begin{table}[H]
\centering
\caption{Irreducible representations (irreps) and basis functions allowed by the normal-state symmetry group for UTe$_2$, D$_{2h}$ $\times$ T $\times$ U(1). The selection rule in the Josephson coupling between an $s$-wave and UTe$_2$ is shown. $J_s$ is the Josephson current density on the respective junction planes at $T$ = 0.}
\begin{tabular}{ccccc}
\toprule
Irrep & Basis function & $J_s$ (001) & $J_s$ (1-10) & $J_s$ (0-11) \\
\midrule
$\Gamma_1^+$ or $A_{1g}$ & $ k_x^2 + k_y^2 + k_z^2$ & $\neq 0$ & $\neq 0$ & $\neq 0$   \\
$\Gamma_2^+$ or $B_{1g}$ & $ k_x k_y$ & $=0$ & $=0$ & $=0$   \\
$\Gamma_3^+$ or $B_{2g}$ & $ k_x k_z$ & $=0$ & $=0$ & $=0$   \\
$\Gamma_4^+$ or $B_{3g}$ & $ k_y k_z$ & $=0$ & $=0$ &$\neq0$   \\
$\Gamma_1^-$ or $A_{1u}$ & $ \hat{x} k_x + \hat{y} k_y + \hat{z} k_z$ & $=0$ & $=0$ & $=0$   \\
$\Gamma_2^-$ or $B_{1u}$ & $ \hat{x} k_y+ \hat{y} k_x$ & $\neq 0$ & $=0$ & $\neq 0$  \\
$\Gamma_3^-$ or $B_{2u}$ & $ \hat{x} k_z+ \hat{z} k_x$ & $=0$ & $\neq 0$ & $\neq 0$  \\
$\Gamma_4^-$ or $B_{3u}$ & $ \hat{y} k_z+ \hat{z} k_y$ & $=0$ & $\neq 0$ & $=0$  \\
\bottomrule
\end{tabular}
\label{Table:1}
\end{table}

Consider now a Josephson junction between an $s$-wave on the left and a non-$s$-wave superconductor on the right of the junction plane featuring a tunnel barrier. The junction orientation is specified by the unit vector $\hat{n}$ perpendicular to the junction plane pointing to the right. The Josephson current density between two spin-singlet superconductors at a fixed phase difference is given by\cite{MRS_semiclassicalJC_1988}

\begin{equation}
\label{eq:JC_S}
J_s \sim \langle (\hat{n} \cdot \hat{k}) Im{ [\Delta_s \Delta^*(\hat k) ]} \rangle _{FS},
\end{equation}

where $\Delta_s$ is the OP of the $s$-wave superconductor, $\Delta(\hat k)$ is the OP of the non-$s$-wave, $\hat k$ is the quasimomentum vector normalized by the Fermi momentum, and $\langle ...\rangle _{FS}$ represents an integral over the Fermi surfaces (FS) that are different for the two superconductors, taking into account also the square of the tunneling matrix element. For a similar junction between an $s$-wave and odd-parity, spin-triplet superconductor, whose Josephson coupling is facilitated by SOC, the corresponding formula is given by\cite{GeshkenbeinLarkin_1986,MRS_semiclassicalJC_1988} 

\begin{equation}
\label{eq:JC_T}
J_s \sim Im{\langle \Delta_s^*(\hat k) \vec d(\hat k) \cdot (\hat k\times \hat{n}) \rangle _{FS}},
\end{equation}
 
where $\vec d(\hat k)$ is the un-normalized order parameter of the odd-parity superconductor on the right of the junction plane. Based on these two equations, expectations on whether J$_s$ will vanish in junctions along certain crystalline orientations can be obtained, which is known as the "selection rule" for Josephson coupling. These expectations for the (001), (1-10), and (0-11) surfaces of UTe$_2$ are shown in Table 1. Incidentally, the Josephson effect experiment was one of the first to show that Sr$_2$RuO$_4$ was most likely a spin-triplet superconductor\cite{JinLiu_2000} before the more decisive phase sensitive experiment was performed\cite{Nelson_GLB_SQUID_2004,LeggettLiu_2021}.

The selection rule in the Josephson coupling requires that the coupling is of the first order, which can in practice be inferred from the zero-temperature limit of I$_c$R$_N$, where I$_c$ is the critical current and R$_N$ is the normal-state junction resistance, respectively. For an all-$s$-wave Josephson junction, it was shown that I$_c$ in the limit of zero temperature was given by the Ambegaokar-Baratoff (A-B) formula\cite{AB_1963} derived from the microscopic tunneling theory assuming, correctly, that the single-electron and pair tunnelings are governed by the same matrix elements, leading to

\begin{equation}
\label{eq:AB}
I_cR_N = \frac{1}{e} \Delta_1 K{[1-(\Delta_1/\Delta_2)^2]^{1/2}},
\end{equation}

where $\Delta_1 $ and $\Delta_2 $ are the superconducting energy gaps of the two $s$-wave superconductors. For a Josephson junction of an $s$-wave and odd-parity superconductor, I$_c$R$_N$ is expected to be small even for the first-order Josephson coupling, as shown theoretically\cite{Tanaka_s-wave_SRO_2003}. If the coupling were of second order, I$_c$R$_N$ would be orders of magnitude smaller than that calculated using Eq. $\ref{eq:AB}$ (for a junction of two $s$-wave superconductors featuring two T$_c$ values that are the same as those used in the present study).

We prepared Josephson junctions of In, an $s$-wave superconductor, and UTe$_2$ by pressing an In dot onto a flat surface of the UTe$_2$ crystal formed naturally during the crystal growth. All crystals used in this study are from the same growth batch. The junctions are shown schematically in Fig. $\ref{fig:1}$C. Given that the preparation of the junction took place in air, an insulating layer of InO$_x$ must form between In and UTe$_2$, which functions as the tunneling barrier. The orientation of the surface was determined by Laue diffraction after all low temperature measurements were completed. Results of sample resistance ($R$) as a function of temperature, $T$, obtained on three junctions prepared on two single crystals of UTe$_2$ are shown in Fig. $\ref{fig:1}$D. The presence of zero resistance indicates that the Josephson coupling is finite. 

\begin{figure}[H]
\centering
\includegraphics[scale=1]{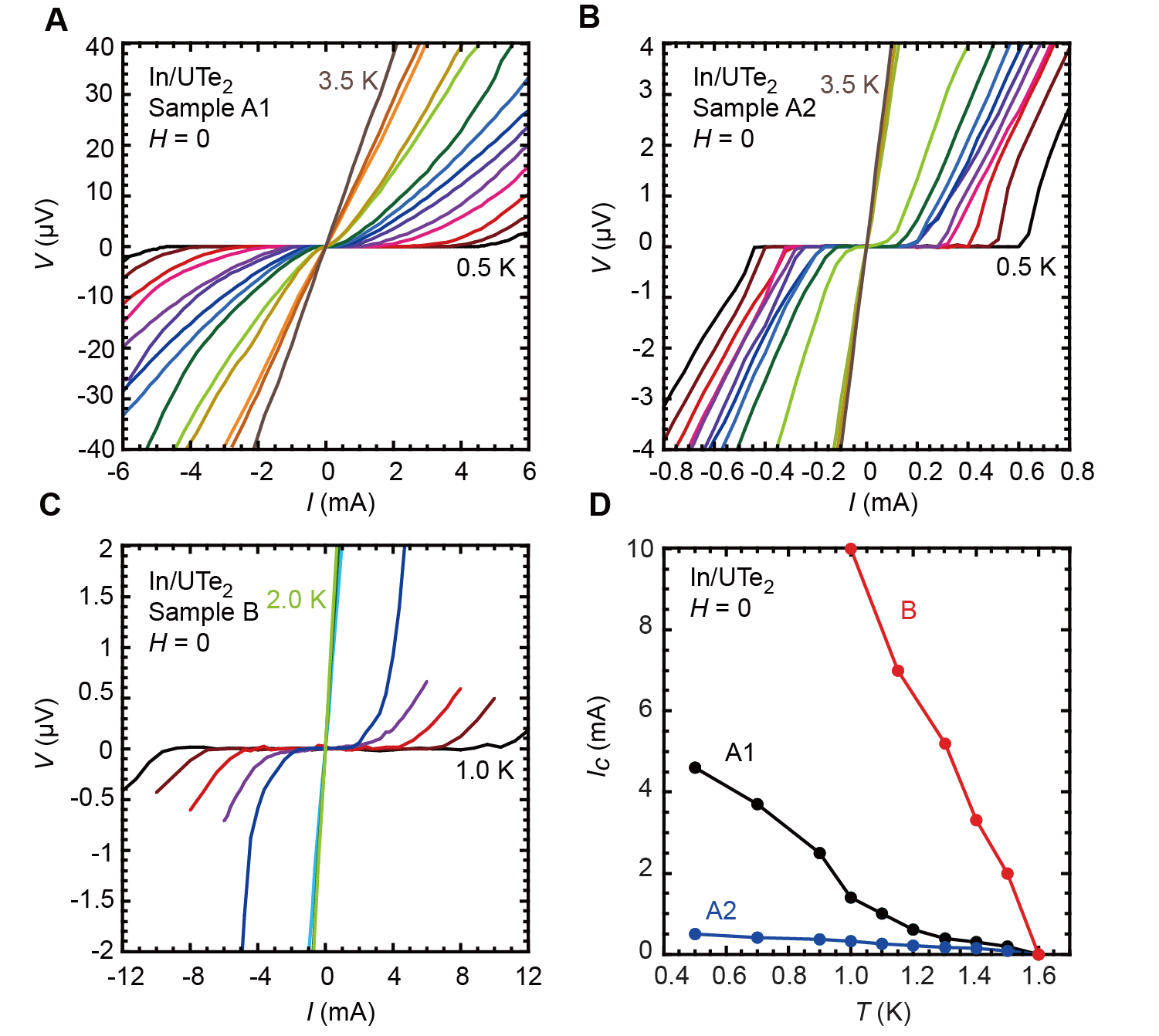}
\phantomcaption
\label{fig:2}
\end{figure}
\noindent {\bf Fig. 2. Finite Josephson coupling in In-UTe$_2$ junctions prepared on (001) surface.} 
(A-C) Voltage ($V$) $vs$. current ($I$) curves for Sample A1 at a fixed temperature $T$ = 0.5, 0.7, 0.9, 1.0, 1.1, 1.2, 1.3, 1.4, 1.5, 1.6, 1.7, 2.5, 3.0, and 3.5 K, Sample A2 at $T$ =  0.5, 0.7, 0.9, 1.0, 1.1, 1.2, 1.3, 1.4, 1.5, 1.6, 2.5, 3.0, and 3.5 K, and Sample B at $T$ = 1.0, 1.15, 1.3, 1.4, 1.5, 1.6, 1.8, and 2.0 K,respectively showing the presence of finite Josephson coupling. For Sample B, the critical current, I$_c$, cannot be measured below 1.0 K due to the heating effect. Junctions A1 and A2 were prepared on the same crystal surface. The Josephson coupling vanishes at T$_c \approx$ 1.6 K. (D) I$_c$ ($T$) for Samples A1, A2, and B. 

Curves of the tunneling current, $I$, $vs$. the bias voltage, $V$, obtained on Samples A1, A2 and B are shown in Fig. $\ref{fig:2}$A-C, revealing a finite critical current (I$_c$) for these junctions. The temperature dependence of I$_c$ for individual junctions is shown in Fig. $\ref{fig:2}$D. The qualitative behavior of I$_c$(T) is as expected even though its precise functional form is yet to be determined theoretically. A kink is seen in I$_c$ (T) for Sample A1, which may be due to the presence of two parallel Josephson junctions with different values of T$_c$. 

\begin{figure}[H]
\centering
\includegraphics[scale=1]{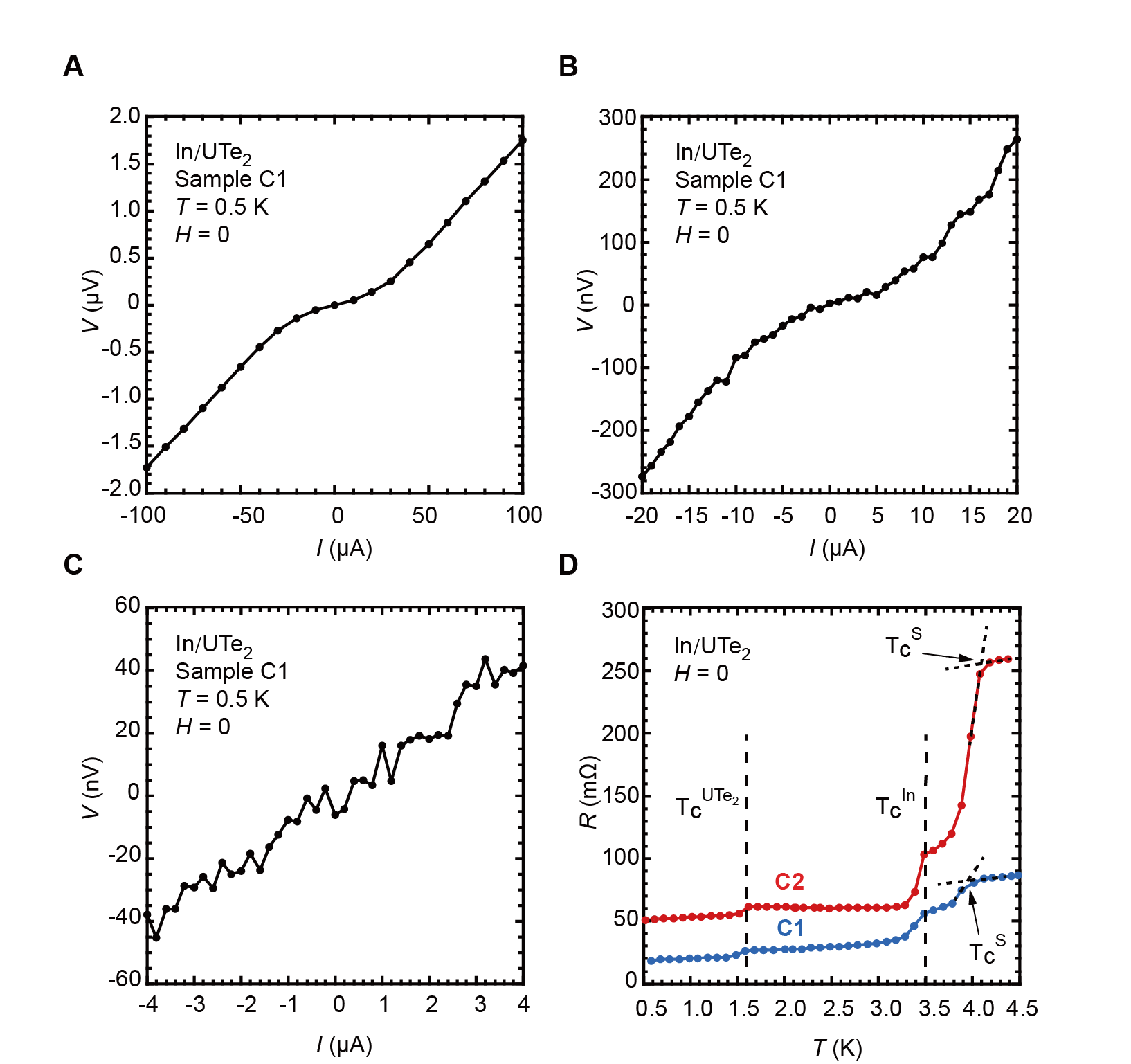}
\phantomcaption
\label{fig:3}
\end{figure}
\noindent {\bf Fig. 3. Zero Josephson coupling in In-UTe$_2$ junctions prepared on (1-10) surface.} 
(A-C) $V$-$I$ curves for Sample C$_1$ at $T$ = 0.5 K measured in decreasing ranges of current. No zero-voltage currents were seen down to the smallest measurement currents. (D) R($T$) curves in zero magnetic fields for Samples C1 and C2. Resistance drops at 1.6 and 3.4 K correspond to the T$_c$'s for UTe$_2$ and In, respectively. The feature seen near 4 K is discussed in the main text. 

$I-V$ curves obtained from Sample C$_1$ at 0.5 K are shown in Figs. $\ref{fig:3}$A-C. No supercurrents are seen even at the lowest temperature. To ensure that we do not miss a very small supercurrent, we systematically reduced the range of current over which the $I-V$ curve was measured. It is evident that at least down to a measurement current range of 4 $\mu$A, the supercurrent, if it exists at all, must be significantly smaller than, say, 1 $\mu$A, which suggests that the Josephson coupling for this sample must be tiny or zero. The same behavior was found in Sample C$_2$. Can the absence of Josephson coupling be due to a large value of R$_N$, which would make I$_c$ small according to Eq. $\ref{eq:AB}$? In Table 1, key parameters of all five samples are summarized. If the values of I$_c$ for these junctions are indeed correlated with R$_N$ as in the case of $s$-wave Josephson junctions (see Eq. $\ref{eq:AB}$), which is apparently the case, the values of I$_c$ for Samples C1 and C2 appear to be at least one or two orders of magnitude small than expected. Given that all five junctions were prepared at roughly the same time using crystals from the same growth batch, the absence of the Josephson junction must be due to symmetry related reasons. 

\begin{table}[H]
\centering
\caption{Summary of parameters for all samples used in the present study. (I$_c$R$_N$) $_{T=0}^{AB}$ is the Ambegaokar-Baratoff (A-B) limit calculated from R$_N$ assuming that the two superconductors are both $s$-wave.}
\begin{adjustbox}{width=\textwidth}
\begin{tabular}{cccccc}
\toprule
Sample & T$_c^{J}$ & R$_N(m\Omega)$ & I$_c$($T_{min}$)(mA) &  I$_c$($T_{min}$)R$_N$ (mV) & I$_c$(T$_{min}$)R$_N$ /(I$_c$R$_N$)$_{T=0}^{A-B}$\\
\midrule
A1 & 1.5 & 19 & 4.6  & 0.087&15\% \\
A2 & 1.5 & 40 & 0.6  & 0.024&4\%\\
B & 1.5 & 5.4 & 10.0  & 0.054&9\% \\
C1 &  & 84 & 0 & 0&N/A\\
C2 &  & 253 & 0 &0&N/A\\
\bottomrule
\end{tabular}
\label{Table:2}
\end{adjustbox}
\end{table}

The orientation of the junction plane were determined after all low-temperature measurements were completed using Laue diffraction. Given the relative sizes of our X-ray beam and the crystals, the entire surface on which the junction or junctions were made was imaged. The symmetry analyses and simulations aimed at determining the surface orientation suggest that Samples A1, A2, and B were prepared on a (001) surface, which happens to be a cleavage plane reported previously\cite{Jiao_STMSTS_chiral_2020}. However, the orientation of the surface on which Samples C1 and C2 were prepared is a (1-10) surface. Remarkably, the presence of Josephson coupling on a (001) and the absence of it on (1-10) suggest that, among all symmetry-allowed pairing states, the symmetry of OP adopted in UTe$_2$ must br $\Gamma_2^-$, or B$_{1u}$ (see Table \ref{Table:1}).

It has been discussed in the literature that UTe$_2$ may feature "accidental degeneracy" in terms of its pairing symmetry\cite{Jiao_STMSTS_chiral_2020}. In the present case, our selection rule results would be consistent with an OP that features an accidental degeneracy of $\Gamma_2^-$ and another symmetry allowed pairing state with zero Josephson coupling for (1-10) surface. Inspection of Table \ref{Table:1} would then suggest that a pairing state mixing $\Gamma_1^-$ and $\Gamma_2^-$ is the only option given that mixing odd and even parities are not allowed in bulk UTe$_2$ in zero magnetic field because of the presence of the inversion symmetry\cite{LeggettLiu_2021}. 

The selection-rule result in Josephson coupling has strong implications on the superconducting properties of UTe$_2$. In the case that no accidental degeneracy exists, we expect the presence of point nodes at the north and south poles for a three-dimensional (3D) Fermi surface (FS). However, if the FS does not contain any points on the axis of $\vec k$ = (0,0,k$_z$), no nodes are expected. Recent ARPES\cite{MiaoWray_FS_ARPES_2020} and de Haas-van Alphen quantum oscillation measurements on UTe$_2$\cite{Aoki_FS_QO_2022} suggested that none of the $k_x = k_y = 0$ points is on the FS, which would exclude the presence of nodes. However, single-particle tunneling\cite{Jiao_STMSTS_chiral_2020}, specific heat\cite{Aoki_specific_heat_2019} and penetration depth\cite{Metz_PointNode_2019} measurements revealed the presence of large midgap states, suggesting that a 3D FS is present in UTe$_2$ even though it has not been detected thus far.        

Features seen in $R_J$ $vs$. $T$ data from Samples C1 and C2 shown in Fig. \ref{fig:3}D deserve further analysis. While resistance drops seen at 1.6 and 3.4 K clearly corresponds to bulk T$_c$'s of UTe$_2$ and In, respectively, a similar feature seen around 4 K is puzzling. It is known that thin films of In under strain can attain a T$_c$ up to 4.4 K\cite{Garland_Tc_enhancement_1971,Anderson_Penetration_depth_1972}. Our junctions were prepared by pressing an In dot hard onto the crystal. The thermal contractions are different for the two superconductors. As a result, an elevated T$_c$ for In is not unexpected near the interface region of In and UTe$_2$. Such elevated T$_c$ must be sample dependent. 

\begin{figure}[H]
\centering
\includegraphics[scale=1]{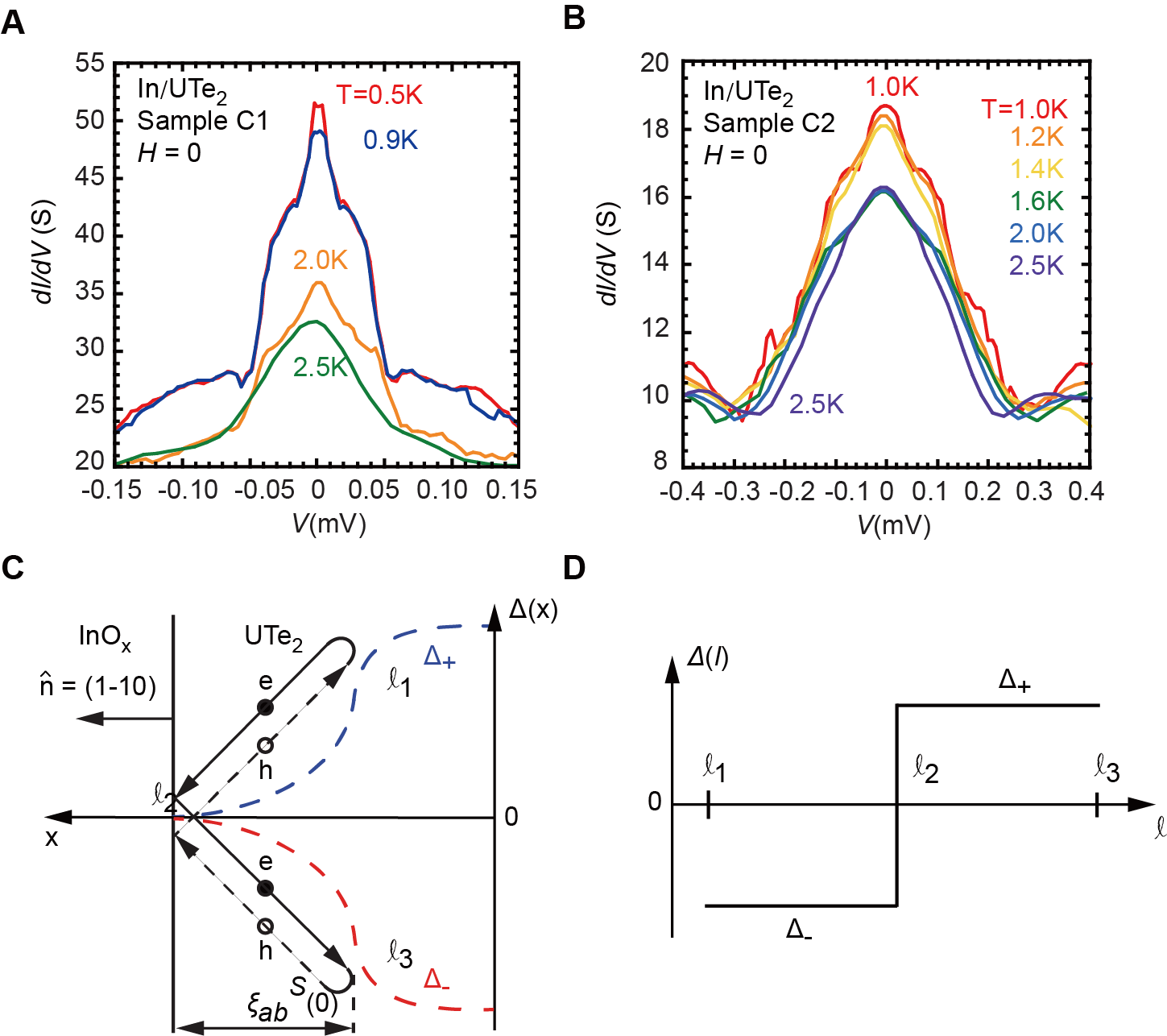}
\phantomcaption
\label{fig:4}
\end{figure}
\noindent {\bf Fig. 4. Formation of Andreev surface bound states.}
(A) Values of d$I$/d$V$ $vs$. $V$ for Sample C1 at $T$ = 0.5, 0.9, 2.0, 2.5 K. (B) Values of d$I$/d$V$ $vs$. $V$ for Sample C2 at $T$ = 1.0, 1.2, 1.4, 1.6, 2.0, 2.5 K. The zero-bias conductance peak is dramatically higher below the T$_c$ of UTe$_2$. (C) Schematic illustrating the formation of Andreev surface bound states (ASBSs). In the classical trajactory for such a state, an electron is Andreev reflected as a hole and reflected off the surface of the crystal as a hole before it is Andreev reflected again as an electron. Here $\Delta_+ (x)$ and $\Delta_- (x)$ are OP for electrons and holes traveling in opposite directions and $\Delta_+ = \Delta_- = 0$ at $l_2$. The superconducting coherence length $\xi_{ab}^S(0)$ is indicated. (D) Schematic illustrating the sign change of OP along the classical trajectory. The superconducting energy gap vanishes at the surface where the OP changes sign. 

Interesting structure in single-particle tunnel spectra, d$I$/d$V$ $vs$. $V$ curves, at low bias voltages obtained from Samples C1 and C2 are present in Figs. \ref{fig:4}A and B. It was proposed previously that the formation of zero-energy Andreev surface bound states (ASBSs) lead to a peak near zero bias voltage in the tunneling spectrum in high-T$_c$ cuprates featuring a pairing state of $\Gamma_3^+$, or B$_{1g}$\cite{hu_MidgapOriginal_1994,yang_MidgapRobustness_1994}, in which the ASBSs were found on the (110) surface\cite{Covington_observation_1997}. Near the (1-10) surface of UTe$_2$, the superconducting OP is expected to be suppressed from the bulk value to zero as the surface is approached as shown in Fig. \ref{fig:4}C if the pairing state is $\Gamma_2^-$. An electron-like quasiparticle with its energy lower than the bulk gap reflected off the surface (junction plane) will be Andreev reflected as it approaches the interior of the crystal (when its energy matches the position dependent gap). The Andreev reflected hole-like quasiparticle will be reflected off the surface and then Andreev reflected again, forming a closed-loop classical trajectory at this particular incident/reflection angle. Importantly, the superconducting OP along this classical trajectory will change sign at the surface (see Figs. \ref{fig:4}C and D), leading to the formation of the ASBSs. It is seen in Figs. \ref{fig:4}A and B that the shape of the conductance peak at low bias voltages changes abruptly across the T$_c$ of UTe$_2$, suggesting that the peak at low bias voltages is associated with the formation of ASBSs. Interestingly, the OP does not change its sign along the classical trajectories formed near the (001) surface of UTe$_2$, suggesting that no ASBSs are expected. Examination of d$I$/d$V$ $vs$. $V$ curves of Samples A1, A2, and C with $V$ sufficiently large indeed suggests that very little feature other than a sharp peak originating from the Josephson coupling is present (data not shown), consistent with the theoretical expectation.    

To summarise, finite Josephson coupling between $s$-wave superconductor In and UTe$_2$ has been observed on (001) junctions but not (1-10) ones. This selection-rule result suggests very strongly that the pairing symmetry in UTe$_2$ at zero magnetic fields is that of $\Gamma_2^-$, or the mixture of $\Gamma_1^-$ and $\Gamma_2^-$, ruling out other symmetry allowed spin-triplet and all spin-singlet pairing states. We also observed a features in the single-particle tunneling spectra that we attribute to the formation of ASBSs on (1-10) surface of UTe$_2$. These results provide a strong foundation for understanding other superconducting properties of UTe$_2$, as well as its mechanism of superconductivity. Whether an applied magnetic field, especially a field as high as 20 T will alter the pairing symmetry in this exotic superconductor is to be determined. 

\section*{Acknowledgments}
{\bf Fundings:} The work done at Penn State was supported by the National Science Foundation (NSF) under grant No. DMR 2312899 as well as Pennsylvania State University. Research at the University of California, San Diego (UC San Diego) was supported by the U.S. Department of Energy, Office of Science, Basic Energy Sciences, under Grant No. DE FG02-04-ER45105 and sponsored in part by the UC San Diego Materials Research Science and Engineering Center (UCSD MRSEC) supported by the NSF (Grant DMR-2011924). R.E.B acknowledges the National High Magnetic Field Laboratory which is supported by the NSF through NSF DMR-1644779 and the State of Florida. The authors acknowledge useful discussions with Tony Leggett, Manfred Sigrist, Johnpierre Paglione, Priscila Rosa and are particularly grateful to Prof. Leggett for reading the manuscript before submission. {\bf Author contributions:} Y.L. conceived and designed the study. M.B.M. led the part of the collaboration involving the preparation of UTe$_2$ single crystals used in the experiments. C.M.M., E.L.W., R.E.B. and M.B.M. synthesized the single crystals and performed the initial characterization. Z.L. prepared samples and carried out the low temperature measurements with additional Laue diffraction measurements and modeling done by C.M.M. and Z.L.. Y.L. and Z.L. analyzed the data and prepared early versions of the manuscript with all authors contributing to its completion. {\bf Competing interests:} The authors declare no competing interests. {\bf Data and materials availability:} Data are available upon request.

on UTe$_2$ crystals from the same growth batch (i.e. grown under the same conditions) as those measured in the present experiment. Data was compared against a reference spectra from the Inorganic Crystal Structure Database (ICSD)\cite{boehme_pXRD_1992}. All peaks, which are indexed according to the ICSD, are positioned well with the UTe$_2$ reference data with no impurity peaks detected.  The small amorphous background at lower angles arises from the quartz powder holder.

\clearpage





\bibliography{UTe2}

\bibliographystyle{Science}

\clearpage
 \end{document}